\pdfoutput=1
\documentclass[12pt]{article}
\usepackage{geometry}
\usepackage{graphicx}
\usepackage{natbib}
\usepackage{setspace}
\usepackage{amsmath,amssymb,bbm}

\newcommand{\indicator}[1]{\mathbbm{1}_{\{ {#1} \} }}

\title{An attraction-repulsion point process model for respiratory syncytial virus infections}
\author{Joshua Goldstein, Murali Haran, Ivan Simeonov, \\
John Fricks and Francesca Chiaromonte \\
Department of Statistics, The Pennsylvania State University \\
{\it email:} mharan@stat.psu.edu}
\date{}

\begin{document}
\maketitle{}

\begin{abstract}
How is the progression of a virus influenced by properties intrinsic to individual cells? We address this question by studying the susceptibility of cells infected with two strains of the human respiratory syncytial virus (RSV-A and RSV-B) in an {\it in vitro} experiment. Spatial patterns of infected cells give us insight into how local conditions influence susceptibility to the virus. We observe a complicated attraction and repulsion behavior, a tendency for infected cells to lump together or remain apart. We develop a new spatial point process model to describe this behavior. Inference on spatial point processes is difficult because the likelihood functions of these models contain intractable normalizing constants; we adapt an MCMC algorithm called double Metropolis-Hastings to overcome this computational challenge. Our methods are computationally efficient even for large point patterns consisting of over 10,000 points. We illustrate the application of our model and inferential approach to simulated data examples and fit our model to various RSV experiments. Because our model parameters are easy to interpret, we are able to draw meaningful scientific conclusions from the fitted models.
\end{abstract}

\section{Introduction}\label{sec:intro}
Biologists are interested in studying viral infections and their effects on living organisms. Of interest is the progression of a virus infection, which is a dynamic process influenced by host defense and resources. We can study how properties intrinsic to individual cells affect the susceptibility of cells to become infected. Studying patterns of infection in cell cultures can give us valuable insight into the role differential susceptibility plays in the outcome of viral infections.

We develop a novel spatial point process model to study the susceptibility of cells to infection by one or more strains of a virus. Our parametric approach allows us to infer parameters that describe spatial structure among the infected cells, thereby providing a methodology for studying spatial patterns of infections under variable experimental conditions and at different stages in the progression of the infection. The computational challenges posed by the spatial point process model are considerable; developing a tractable computational Markov chain Monte Carlo approach for Bayesian inference is therefore an important component of this work.

The data we utilize are generated by {\it in vitro} experiments that examine the response of human epithelial cells to infections with the human respiratory syncytial virus (RSV) \citep{simeonov2010exploratory}. RSV is a major cause of respiratory illness and has been classified into strains based on antigenic and sequence data. The strains RSV-A and RSV-B are the focus of the study we consider.

Cell properties can be gleaned from inferences on the spatial structure of infections in our cell cultures. Since the data collected are images which are pre-processed to identify the location of infected cells, this data lends itself to a point process framework. As in the case of many spatial point process models, the model we develop has an intractable normalizing constant which presents a computational challenge. We therefore adopt the {\em double Metropolis-Hastings algorithm} of \citet{liang2010double}, an auxiliary variable algorithm that uses two nested MCMC samplers to sample from distributions with intractable normalizing constants.

Inference on our model suggests the existence of a significant structure in the spatial patterns of cells infected with RSV. We can infer that when cells in close proximity with one another tend to be infected, then there is evidence of susceptibility in these nearby cells. This implies that cells near one another have a similar level of susceptibility and there is some spatial synchronicity in susceptibility states.

Our approach has substantial advantages over simpler nonparametric approaches. Fitting an explicit probability model allows us to simulate processes consistent with the data and to study the sensitivity of the model to changing parameter values. Moreover, parameters in our model formulation lend themselves to direct interpretation and are subject to rigorous inference via 95\% posterior credible intervals -- whereas nonparametric approaches (e.g. fitting curves to pair correlation functions) necessarily rely on ad-hoc techniques to draw conclusions about the characteristics of the spatial structure.

The remainder of this paper is organized as follows. In Section \ref{sec:setup}, we explain how the RSV data were generated and describe our exploratory analysis of the resulting data, which motivated and informed the development of our novel attraction-repulsion point process model. In Section \ref{sec:spp}, we describe our model, the computational challenges it poses and our inferential approach. In Section \ref{sec:application}, we apply our methods to both simulated and RSV data, and draw meaningful scientific conclusions from the latter. We discuss these results and possible future extensions in Section \ref{sec:discussion}.

\section{RSV Data and Exploratory Analysis}\label{sec:setup}
In this section we outline the design of the RSV experiments in Simeonov {\em et al.} (2010) and summarize our exploratory analysis of the resulting data.

\subsection{Experimental Design}
We now describe the experiments conducted to collect the {\em in vitro} RSV data. This description closely follows the one in \cite{simeonov2010exploratory}. Wells are plated with $1 \times 10^5$ bronchial epithelial cells and exposed to one strain of the virus, the ``primary'' strain. These cell cultures are then allowed to develop for either 3 or 16 hours. Next, the cells are exposed to a second strain of the virus, the ``challenge'' strain, which is allowed to act for one hour, after which cells are washed to remove unattached virus. 24 hours after the cells are washed, the locations of cells infected with the primary and challenge strains are obtained through microscopy imaging.

\begin{figure}
\centering
\includegraphics[width = 0.8 \textwidth]{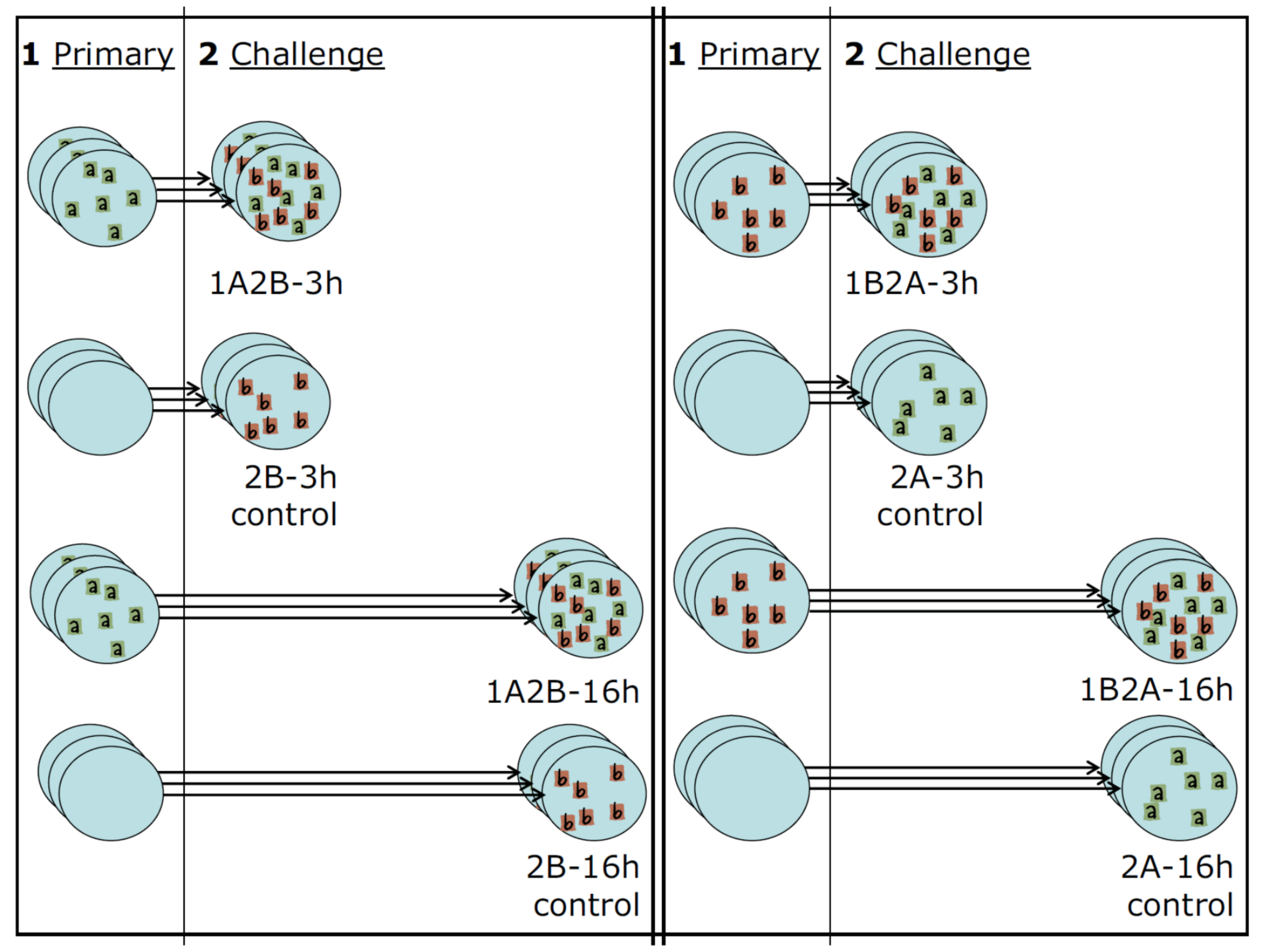}
\caption{Diagram of the experimental setup from \protect\citet{simeonov2010exploratory}.}
\label{fig:experiment}
\end{figure}

When the primary strain is RSV-A and the challenge is RSV-B, we denote the experimental settings by 1A2B-3h and 1A2B-16h. The settings when the order of the strains is reversed are denoted 1B2A-3h and 1B2A-16h. We also consider control settings in which the cultures are {\em not} exposed to a primary strain, but only to a challenge strain after either 3 or 16 hours; the protocol followed is otherwise the same. These settings are denoted by 2A-3h and 2A-16h when RSV-A is the challenge, and 2B-3h and 2B-16h when RSV-B is the challenge. The eight settings above are summarized in Figure \ref{fig:experiment}.

Also of interest are four additional control settings where there is a primary infection but no challenge. The primary strain is introduced and allowed to act for either 3 or 16 hours before imaging. These settings are denoted by 1A-3h and 1A-16h when RSV-A is the primary strain, 1B-3h and 1B-16h when RSV-B is the primary strain. Each of the twelve experimental settings was independently replicated three times.

Microscopy imaging is based on staining procedures that produce fluorescent marks at the locations of infected cells. Image analysis software is utilized to generate 2D spatial coordinates for each such mark in an image. The staining procedure differs for RSV-A and RSV-B. For the former, a fluorescent mark appears throughout the cytoplasm of an infected cell, while for the latter the mark appears on the cell membrane. RSV-A marks are larger and fuzzier. Notably, the staining procedure for RSV-A implies that no two RSV-A marks can be too close together due to cell volume exclusion while there is no such restriction for RSV-B marks. This difference, which creates an artifact when investigating attraction-repulsion behavior at very small scales, is accounted for in our model as presented in Section \ref{sec:spp}. Separate marks are also produced at the locations of cell nuclei.  For more details on experimental methods see \cite{simeonov2010exploratory}.

\subsection{Exploring the RSV Dataset}
Spatial point processes in a plane are a natural framework for the RSV data. We introduce this framework along with basic notation, examine the pair correlation function as a means of exploratory data analysis, and provide the motivation for a new spatial point process model.

\subsubsection{Spatial point process basics}
\begin{figure}
\centering
\includegraphics[width = 0.5 \textwidth]{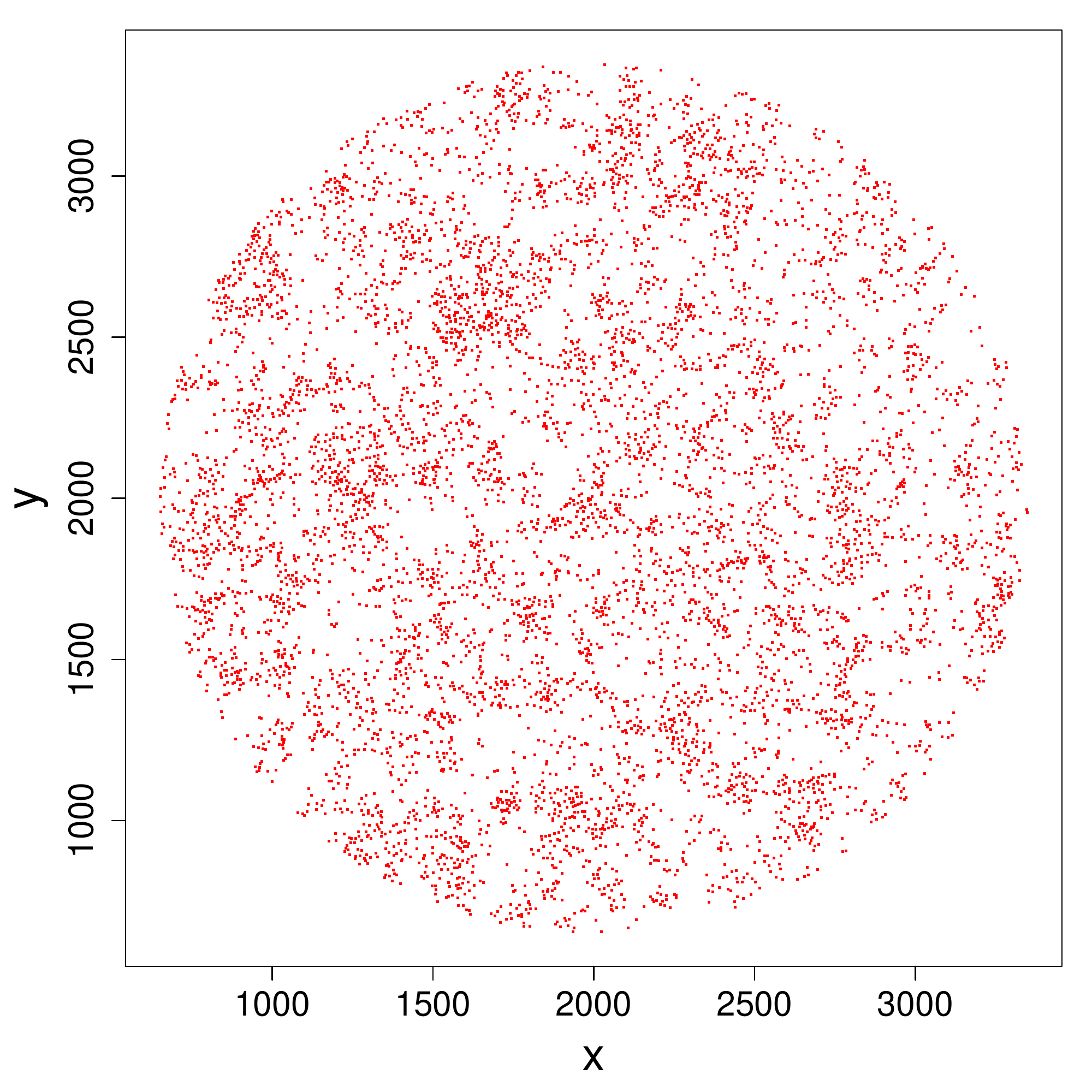}
\caption{Example of point data within a well. Each well image consists of a circular observation window with a radius of 1350 pixels; 1 pixel corresponds to 6.45 microns. Points represent locations of cells infected with RSV.}
\label{fig:sampledata}
\end{figure}

A spatial point process can be thought of as the probabilistic mechanism generating point data, i.e.~a collection of discrete random points in a plane. An instance is provided in Figure \ref{fig:sampledata}, where each point represents the two-dimensional coordinates of an RSV infected cell in a well, as identified by imaging software. One pixel in the microscope image corresponds to 6.45 microns, while cell diameters are on the order of 3-4.5 pixels (19.3-29.0 microns). Here the goal is to characterize the infection status of cells as a function of their location and proximity to one another -- detecting spatial association in the form of \emph{attraction} or \emph{repulsion} among infected cells, i.e. the tendency of cells surrounding an infected cell to have higher or lower susceptibility to infection.

Let $X=(x_1,...,x_n)$ in a bounded region $W \subseteq \mathbb{R}^2$ (the well) indicate the locations of infected cells, and assume $X$ is a realization of a spatial point process where the number of points $n$ is itself random (see e.g. \citet{diggle1983statistical,ripley1991statistical,moller2003statistical}). To capture interactions among points in $X$ we consider whether they are more or less likely to be separated by a distance $r$ than expected under a null, no interaction scenario. When they are more likely we say there is {\it attraction}, and when they are less likely we say there is {\it repulsion} on the scale of $r$. Spatial point process models have been developed to formalize such interactions by explicitly taking into account information about pairs of points. For instance, consider the class of models
\begin{equation}
f(X|\Theta) = \dfrac{ 1} { c(\Theta) } \lambda^n \displaystyle\prod_{i \neq j} \phi(x_i,x_j)
\label{eq:framework}
\end{equation}
obtained for various specifications of $\phi(x_i,x_j)$, which is known as the interaction function between the $i$th and $j$th points. In the stationary case $\phi(x_i,x_j) = \phi(\|x_i-x_j\|)=\phi(r)$, where $r$ is the distance between the two points. A value $\phi(r) > 1$ indicates attraction, a tendency for points to cluster at distance $r$. Similarly, $\phi(r) < 1$ indicates repulsion at distance $r$. The intensity of the process is controlled by $\lambda$, and $c(\Theta)$ is the normalizing constant where $\Theta = (\lambda, \Xi)$, $\Xi$ comprising the parameters of the interaction function. As we will discuss later, the fact that $c(\Theta)$ is typically computationally intractable and depends on $\lambda$ presents serious inferential challenges.

\subsubsection{Pair correlation function}
The pair correlation function (PCF) is a useful means to explore point data generated by stationary spatial point processes. One convenient definition of the PCF is in terms of Ripley's K function. For our collection of points $X=(x_1,...,x_n)$, let $A$ indicate the area of $W$ and $\rho = \frac{n}{A}$ the overall density of points in the region.

Ripley's K is defined as
\[ K(r) = \dfrac{1}{A \rho^2} \text{E} \left[ \displaystyle\sum_{i=1}^n \sum_{j \neq i} \indicator{||x_i - x_j|| < r} \right] \]
$\rho K(r)$ is the expected number of points in a circle of radius $r$ around a typical point in the process, and $K(r)$ the expected density of points in the circle. The PCF is then defined as
\[ g(r) = \dfrac{1}{2\pi r} K'(r) \]
where the derivative $K'(r)$ can be thought of as the change in expected density at radius $r$. For a Poisson process, where the locations of all points are independent, $K(r) = \pi r^2$ and $g(r) = 1$. Values of $g(r) > 1$ or $g(r) < 1$ (changes in expected density higher or lower than if the points were independent) indicate attraction or repulsion at distance $r$. This is analagous to the interaction function above; the PCF gives us the empirical attraction and repulsion behavior, while the interaction function is a parametric specification of this behavior in the model.

There are a number of methods for estimating PCF (see \citet{kerscher2000comparison} for a review). For instance, we can estimate Ripley's K by
\[ \hat{K}(r) = \dfrac{A}{n(n-1)} \displaystyle\sum_{i=1}^n \sum_{j \neq i} \indicator{||x_i - x_j|| < r} e_{i,j} \]
 where $e_{i,j}$ is an edge correction term to account for points close to the boundary of the region. Then an estimate of $g(r)$ can be constructed in terms of the derivative of a smoothed $\hat{K}(r)$.

When performing PCF estimation for exploratory analysis and for assessing goodness of fit, we adopt the bootstrap method of \citet{loh2008valid}. The advantage of this method is that we can easily compute confidence intervals of our estimate for a single observed point pattern with no replicates. To find the estimate, we first compute an array of $n$ local pair correlation functions for each point in the pattern. The local PCF, also called the local indicator of spatial association (LISA) is the contribution to the PCF from each data point. The local PCF of point $i$ is computed by kernel smoothing as in \citet{stoyan1994fractals}:
\[ \hat{g}_i(r) = \dfrac{A}{2 \pi n r} \displaystyle\sum_{j \neq i} k(||x_i-x_j|| - r) \]
\[ k(t) = \dfrac{3}{4 \delta}\text{max}\left(0, 1 - \dfrac{t^2}{\delta^2}\right) \]
where $\delta$ is a smoothing parameter. The PCF is then estimated by $\hat{g}(r) = \frac{1}{n}\sum_{i=1}^n \hat{g}_i(r)$. To get a bootstrap estimate, let $N^*$ be a resampling of $n$ numbers from $(1,...,n)$ with replacement. Then $\frac{1}{n}\sum_{i\in N^*} \hat{g}_i(r)$ becomes one bootstrap estimate. This is repeated $B$ times to get a bootstrap sample. Intervals are constructed by taking pointwise 95\% quantiles of the bootstrap sample. For the RSV data, local PCFs are sampled with equal probability from each of the three replicates.

\subsubsection{Motivation for a new model}
\begin{figure}
\centering
\includegraphics[width = \textwidth]{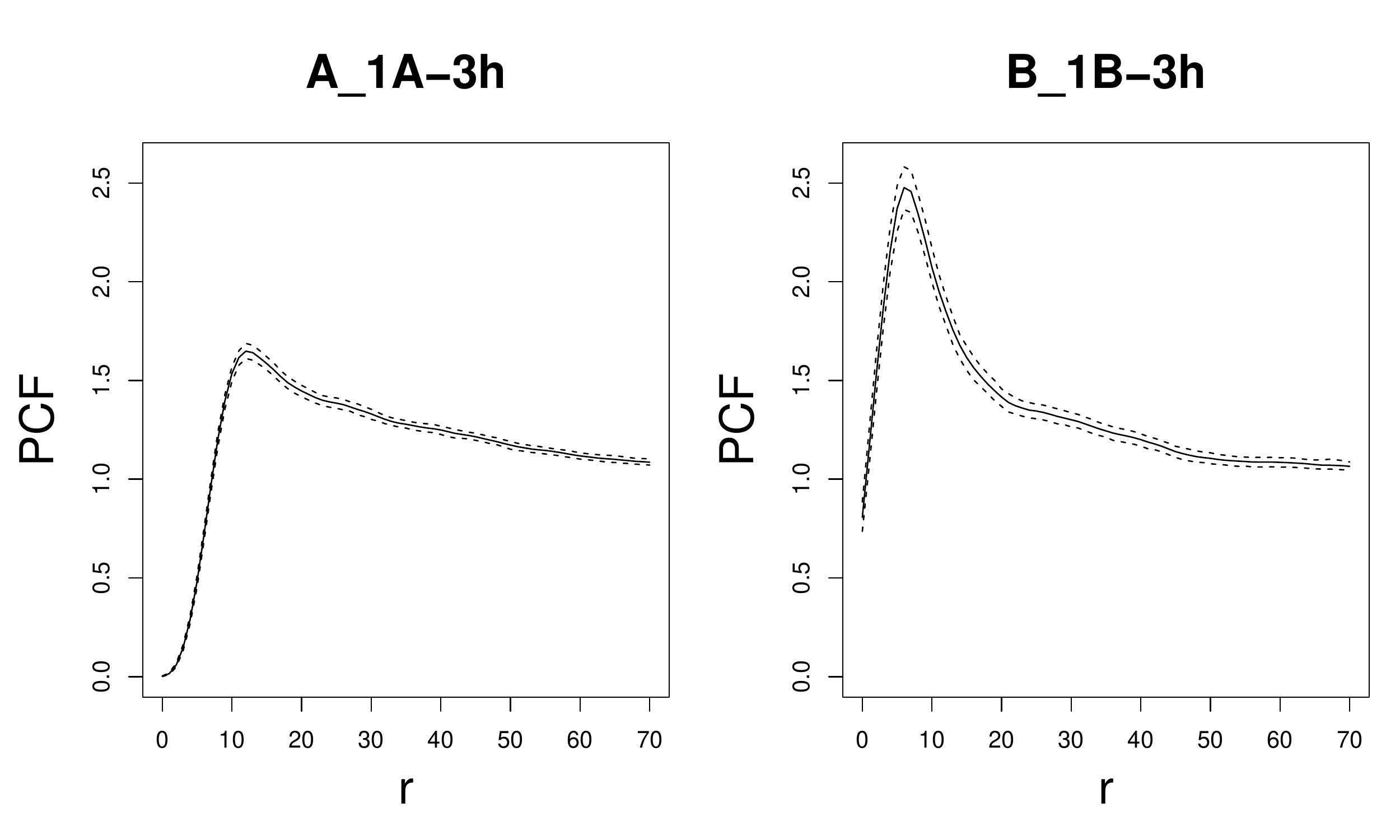}
\caption{Pair correlation functions of RSV-A (left) and RSV-B (right) estimated from the $A_{1A-3h}$ and $B_{1B-3h}$ experiments. Estimates were obtained using the bootstrap method of \protect\citet{loh2008valid}. Values of PCF$(r) > 1$ indicate attraction between points on the scale of $r$, while PCF$(r) < 1$ indicates repulsion. There is a small-scale repulsion due to cell volume exclusion. This is very pronounced for RSV-A, where infection marks appear in the cytoplasm and thus cannot be too close together, and less so for RSV-B, where infection marks appear on cell membranes. Attraction at moderate $r$ indicates an increased susceptibility to infection for cells near infected cells. Notably, this is much stronger for RSV-B than for RSV-A, suggesting a possible difference in the cell responses elicited by the two strains of the virus.}
\label{fig:samplePCF}
\end{figure}

A spatial point process model for the RSV data must be able to capture attraction-repulsion behavior indicative of the tendency of infections to lump together or remain apart at various spatial scales. The PCF gives us a means to explore such behavior. Figure $\ref{fig:samplePCF}$ illustrates the general features of PCFs estimated from the RSV data. We observe repulsion at small $r$ (PCF below $1$), a peak of attraction at moderate $r$ (PCF above $1$) and then declining evidence of association as $r$ increases (PCF falling towards $1$). In other words, infection marks are less likely than expected to be very close together, cluster together more than expected at intermediate scales, and become roughly independent at sufficiently large scales. This suggests that the attraction-repulsion behavior of the underlying process is a highly structured, smoothly-varying function of the scale $r$. With this information at hand, our aim is to develop a spatial point process model cable of explicitly and parametrically capturing such a behavior.

\section{A New Spatial Point Process Model for Attraction-Repulsion}\label{sec:spp}
In this section, we describe our novel spatial point process model for RSV data and explain its connections to the Strauss process first introduced in \citet{strauss1975model} and related models due to \citet{geyer1999likelihood}. Because there are significant computational challenges to address, we perform inference for our model via the double Metropolis-Hastings algorithm \citep{liang2010double}.

\subsection{Model for RSV data}
A simple model that incorporates repulsion is the Strauss process \citep{strauss1975model,geyer1999likelihood}. This model falls into the framework of (\ref{eq:framework}), with an interaction function given by
\[ \phi(x_i,x_j) = \begin{cases} \gamma, & 0<\|x_i-x_j\| \leq R \\ 1, & \|x_i-x_j\| > R \end{cases} \]
where $0\leq \gamma \leq 1$. Since $\phi(x_i,x_j) \leq 1$ this is a purely repulsion point process. A more sophisticated model due to \citet{geyer1999likelihood} allows for both attraction and repulsion. The likelihood is given by
\[ \mathcal{L}(X|\Theta) = \displaystyle\frac{ \lambda^n \left[ \displaystyle \prod_{i=1}^n \text{exp} \left\{ \gamma \text{ min} \left( \sum_{i \neq j} 1_{ \{ \| x_i - x_j \| \leq R \} }, k \right) \right\} \right]} {c(\Theta)} \]
where $k>0$ and $\gamma \in (-\infty,\infty)$. The strength of the attraction varies by the number of pairs of points within a distance $R$ of each other. Note that the constant $k$ is introduced to prevent attraction from becoming too strong, which can lead to degenerate behavior wherein sample realizations from the model will consist of large clumps of points clustered together. To capture the complex nature of the spatial associations observed in the RSV data, we formulate an extension of this model where a flexible interaction function allows attraction to vary smoothly over distance.
\subsubsection{Interaction function}
We define the interaction function $\phi(r)$ between two points at distance $r$ piecewise, as
\[ \phi(r) = \begin{cases} 0, & 0\leq r \leq R \\
y_1(r) \equiv \theta_1 - \bigg( \dfrac{ \sqrt{\theta_1} }{ \theta_2 - R } (r - \theta_2) \bigg)^2, & R < r \leq r_1 \\
y_2(r) \equiv 1 + \dfrac{ 1 }{ (\theta_3 (r - r_2))^2 }, & r > r_1 \\
\end{cases} \]
Here $\theta_1$ is the value of $\phi(\cdot)$ at its peak, $\theta_2$ is the value of $r$ at the peak, and $\theta_3$ controls the rate of descent after the peak. $R$ is the minimum allowable distance between points, and $r_1$ and $r_2$ are solutions to the system of equations
\[ \begin{cases}
y_1(r_1) = y_2(r_1) \\
\dfrac{dy_1}{dr}(r_1) = \dfrac{dy_2}{dr}(r_1) \\
\end{cases} \]
such that the interaction function is continuously differentiable.

\begin{figure}
\centering
\includegraphics[width = \textwidth]{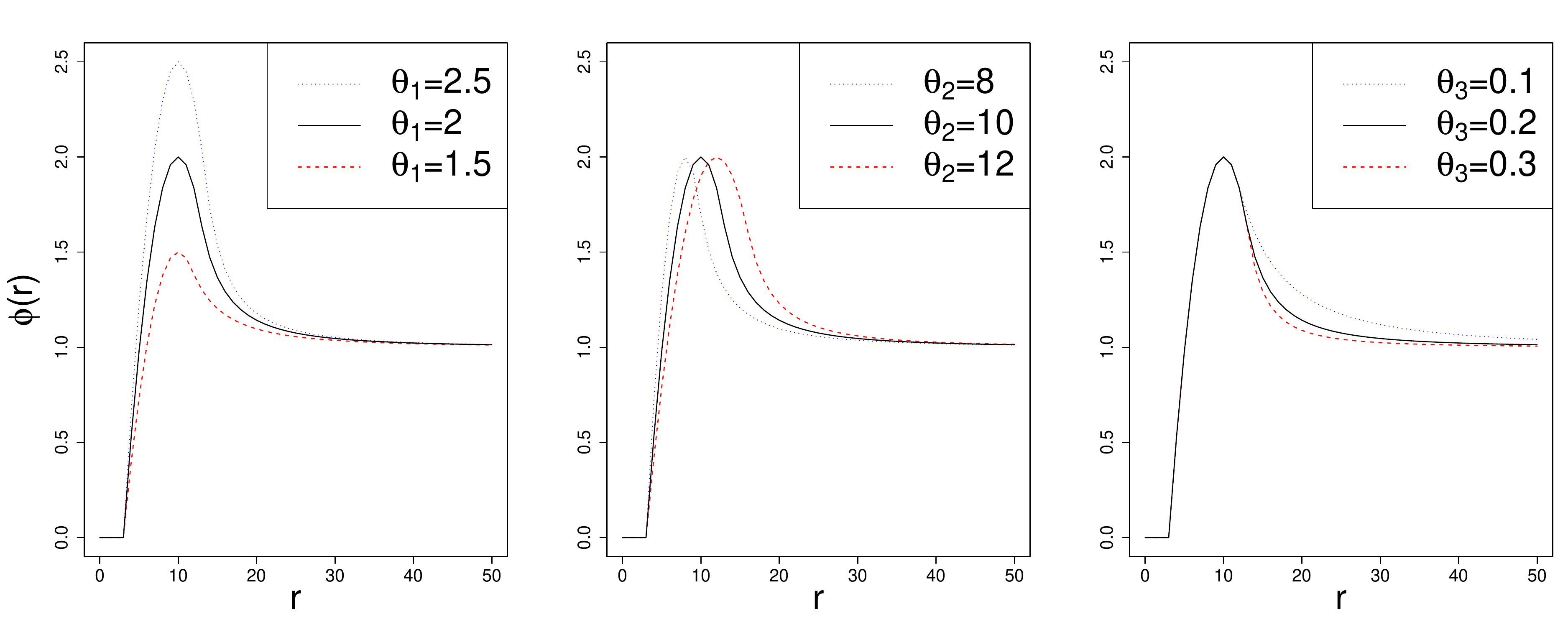}
\caption{Sample plots demonstrating how the interaction function $\phi(r)$ changes with values of the peak height $\theta_1$, the peak location $\theta_2$, and the rate of descent $\theta_3$.}
\label{fig:sampleinteract}
\end{figure}

We saw that the RSV data exhibits a complex attraction-repulsion behavior that varies smoothly with scale. This interaction function can fit all the features we observed in our estimated PCFs (see Figure \ref{fig:samplePCF}): a small scale repulsion followed by a peak attraction and a descent to $1$ at large scales. Using only quadratic functions, it has the flexibility to capture a broad range of shapes of the type illustrated in Figure \ref{fig:sampleinteract} and is restricted to be continuously differentiable to ensure the interaction is smoothly varying with $r$. The incorporation of a minimum allowable distance $R$ lets us capture artifacts of the cell-staining procedures that differ between RSV-A and RSV-B, such as the cell volume exclusion effect discussed in Section \ref{sec:setup}.

\subsubsection{Likelihood}
Let $\Theta = (\lambda, k, \theta_1, \theta_2, \theta_3)$. The likelihood of our model is given by
\[ \mathcal{L}(X|\Theta) = \dfrac{ \lambda^n \left[ \displaystyle \prod_{i=1}^n \text{exp} \left\{ \text{min}\left(\sum_{i \neq j} log(\phi(x_i,x_j)), k \right) \right\} \right] } {c(\Theta) } \]
$\theta_1$, $\theta_2$ and $\theta_3$ determine the shape of the interaction function which tells us at what scales we have attraction and repulsion. The parameter $\lambda$ controls the intensity, $k$ is a truncation constant to prevent degenerate ``clumping" behavior as before, and $R$ is the minimum distance allowed between any two points. In practice, inference on R is problematic. For RSV-A we fix $R$ at the minimum distance between observed marks. This is about 3 pixels (19.3 microns), on the order of the smallest cell diameters. RSV-B experiments do not have the cell volume exclusion problem and we therefore fix $R=0$. Finally, $c(\Theta)$ is an intractable normalizing constant.

\subsection{Inference using double Metropolis-Hastings}
We wish to perform Bayesian inference on the parameters $\Theta$. Because the likelihood contains an intractable normalizing constant, traditional Markov chain Monte Carlo (MCMC) methods cannot be applied. \citet{murray2006} refers to these models as having ``doubly intractable'' distributions. In a frequentist setting, this can be handled by replacing the intractable likelihood with an approximate pseudolikelihood \citep{besag1974spatial}. Maximum likelihood inference is also possible using MCMC-MLE \citep{geyer1992constrained} and has been applied to spatial point processes (e.g. \citet{geyer1999likelihood,moller2003statistical,simeonov2012}). Standard errors for MCMC-MLE are difficult to obtain for our model because analytical gradients of the unnormalized likelihood are not available; this is important for implementation of MCMC-MLE algorithms as described in \cite{geyer1992constrained}. Initializing the algorithm is problematic as well, requiring a grid search over the parameter space.

Recently some efforts have been made to do inference on doubly intractable models in a Bayesian setting. \citet{atchade2008bayesian} proposes an MCMC sampler where the normalizing constant is approximated using the Wang-Landau algorithm \citep{wang2001efficient}. We use an algorithm based on the double Metropolis-Hastings (DMH) algorithm following \citet{liang2010double}.

\subsubsection{Markov chain Monte Carlo}
In a Bayesian setting, given data $X$, parameters $\Theta$ and prior $p(\Theta)$, inference for $\Theta$ is based on the posterior distribution
\[ \pi(\Theta|X) \propto \mathcal{L}(X|\Theta) p(\Theta) \]
MCMC is a common method of Bayesian inference. In MCMC, the goal is to construct a Markov chain that has the target distribution $\pi(\Theta|X)$ as its stationary distribution. The Metropolis-Hastings algorithm is a method of constructing such a chain. At each step of the algorithm, one proposes $\Theta'$ from a proposal function $q(\Theta,\Theta')$ and calculates the following acceptance probability:
\[ \alpha = \text{min} \left(1,\displaystyle\frac{p(\Theta')q(\Theta',\Theta)\mathcal{L}(X|\Theta')}{p(\Theta)q(\Theta,\Theta')\mathcal{L}(X|\Theta)} \right) \]
In the case where the likelihood contains an intractable normalizing constant, we write $\mathcal{L}(X|\Theta) = \dfrac{ h(X|\Theta) }{ c(\Theta) }$. The acceptance probability becomes
\[ \alpha = \text{min} \left(1,\displaystyle\frac{p(\Theta')q(\Theta',\Theta)h(X|\Theta')}{p(\Theta)q(\Theta,\Theta')h(X|\Theta)} \times \frac{c(\Theta)}{c(\Theta')} \right) \]
The intractable normalizing constant $c(\Theta)$ does not cancel out of the acceptance probability. This traditional method of MCMC cannot be applied. We can get around this dilemma by introducing an auxiliary variable.

\subsubsection{Auxiliary variable algorithm}
The auxiliary variable algorithm \citep{moller2004efficient,moller2006efficient} or exchange algorithm \citep{murray2006} is an ``exact'' method that relies on perfect sampling. The algorithm proceeds as follows: 
\begin{enumerate}
\item If $\Theta$ indicates the current state, propose a transition to $\Theta'$ from $q(\Theta,\Theta')$.
\item Generate an auxiliary variable $Y \sim \displaystyle\frac{1}{c(\Theta')}{h(Y,\Theta'})$ from a perfect sampler.
\item Accept $\Theta'$ with probability
\[ \alpha = \text{min} \left(1,\displaystyle\frac{p(\Theta')q(\Theta',\Theta)h(X,\Theta')h(Y,\Theta)}{p(\Theta)q(\Theta,\Theta')h(X,\Theta)h(Y,\Theta')} \times \frac{c(\Theta)c(\Theta')}{c(\Theta')c(\Theta)} \right)
\]
\end{enumerate}
The normalizing constants cancel.

\subsubsection{Double Metropolis-Hastings}
The double Metropolis-Hastings (DMH) algorithm of \citet{liang2010double} is an approximate version of the exchange algorithm. Instead of generating the auxiliary variable $Y$ from a perfect sampler, we approximate samples from $\frac{1}{c(\Theta')}{h(Y,\Theta'})$ by introducing a second Markov chain for $Y$. The algorithm proceeds as follows:
\begin{enumerate}
\item If $\Theta$ indicates the current state, propose a transition to $\Theta'$ from $q(\Theta,\Theta')$.
\item Beginning at state $X$, generate an auxiliary variable $Y$ through $m$ Metropolis-Hastings updates from a kernel with stationary distribution $\displaystyle\frac{1}{c(\Theta')}{h(Y,\Theta'})$.
\item Accept $\Theta'$ with probability
\[ \alpha = \text{min} \left( 1,\displaystyle\frac{p(\Theta')q(\Theta',\Theta)h(X,\Theta')h(Y,\Theta)}{p(\Theta)q(\Theta,\Theta')h(X,\Theta)h(Y,\Theta')} \right)
\]
\end{enumerate}
The DMH algorithm uses two nested MCMC samplers, with the ``inner'' sampler generating an auxiliary point pattern at each step of the ``outer'' sampler. A greater number of updates $m$ in the inner sampler means that $Y$ comes from a distribution closer to the target distribution $\frac{1}{c(\Theta')}{h(Y,\Theta'})$.

Thus far we have described inference on a single point pattern $X$. Given three independent replicates $\mathbf{X}=(X_1,X_2,X_3)$ as we have for the RSV data, the likelihood function is
\[
\mathcal{L}(\mathbf{X}|\Theta) = \prod_{i=1}^3 \mathcal{L}(X_i|\Theta) = \dfrac{ \prod_{i=1}^3 h(X_i|\Theta) }{ c(\Theta) }
\]
The DMH algorithm proceeds analogously to the above; at each iteration of the outer sampler we generate auxiliary variable $\mathbf{Y}=(Y_1,Y_2,Y_3)$ and accept $\Theta'$ with probability
\[ \alpha = \text{min} \left( 1,\displaystyle\frac{p(\Theta')q(\Theta',\Theta)\mathcal{L}(\mathbf{X},\Theta')\mathcal{L}(\mathbf{Y},\Theta)}{p(\Theta)q(\Theta,\Theta')\mathcal{L}(\mathbf{X},\Theta)\mathcal{L}(\mathbf{Y},\Theta')} \right) =
\text{min} \left( 1,\displaystyle\frac{p(\Theta')q(\Theta',\Theta)}{p(\Theta)q(\Theta,\Theta')}\prod_{i=1}^3 \frac{h(X_i,\Theta')h(Y_i,\Theta)}{h(X_i,\Theta)h(Y_i,\Theta')} \right)
\]

\subsubsection{The inner sampler}
For our spatial point process model, the inner sampler is a birth-death MCMC sampler. At each step of the chain, we propose to either remove an existing point or add a new point with equal probability. If we propose to remove a point, the point is selected uniformly at random from the current $n$ points. If we propose to add a point, the new point is selected uniformly over the state space. In either case we then calculate the appropriate acceptance probability.

To formalize this algorithm, again let $X = (x_1,...,x_n)$ be a point pattern observed in $W \subset \mathbb{R}^2$ with area $A$. Define
\[ X^+ \equiv X \cup \{ \xi \} \]
be the point pattern formed by adding a new point $\xi$, and
\[ X^- \equiv X \backslash \{ \xi \} \]
be the point pattern formed by deleting an existing point $\xi \in \{x_1,...,x_n\}$. At each iteration, we choose to add a point with probability $p_1$ or delete an existing point with probability $p_2 = 1 - p_1$. Points are added at a location according to the density $b(X, \xi) = \frac{1}{A}$, that is, points are added uniformly over the region $W$. A point is selected for deletion by drawing from $d(X, \xi) = \frac{1}{n(X)}$, that is, points are selected for deletion with equal probability from the set $X$. Assume we are in state $\Theta$ in the ``outer'' sampler and our goal is to generate a sample from $\frac{1}{c(\Theta')}{h(Y,\Theta'})$. Given a point pattern $X_t$ at iteration $t$, the algorithm proceeds as follows:
\begin{itemize}
\item Simulate $U \sim \text{Unif}(0,1)$. If $U < p_1$, add a point:
\begin{enumerate}
\item Generate the proposal point pattern $X^+$ by drawing a new point $\xi$ from $b(X_t, \xi)$.
\item Simulate $V \sim \text{Unif}(0,1)$. If $V \leq \text{min} \left( 1, \displaystyle\frac{ p_2 h(X^+,\Theta) d(X^+,\xi)}{ p_1 h(X_t,\Theta) b(X_t,\xi) } \right)$ set $X_{t+1} = X^+$; else set $X_{t+1} = X_t$.
\end{enumerate}
\item Else if $U > p_1$, delete a point:
\begin{enumerate}
\item Generate the proposal point pattern $X^-$ by drawing an existing point $\xi$ to delete from $d(X_t, \xi)$.
\item Simulate $V \sim \text{Unif}(0,1)$. If $V \leq \text{min} \left( 1, \displaystyle\frac{ p_1 h(X^-,\Theta) b(X^-,\xi)}{p_2 h(X_t,\Theta) d(X_t,\xi) } \right)$ set $X_{t+1} = X^-$; else set $X_{t+1} = X_t$.
\end{enumerate}
\end{itemize}

\subsubsection{Computational challenges}
Since the DMH algorithm uses nested MCMC samplers, it is computationally expensive; the inner sampler must be run for thousands of iterations at each step of the outer sampler. Inner sampler updates are fast in practice since we only need to consider the {\it change} in likelihood caused by adding or removing a single point; this is a consequence of the birth-death sampler. Even so, these likelihood evaluations are a computational bottleneck. However, on the RSV data the interaction function $\phi(r)$ tends to $1$ as $r$ becomes large, so significant computational savings can be obtained by truncating $\phi(r)$, setting it to $1$ when $r > r_{\text{max}}$. The choice of $r_{\text{max}}$ is of course application specific; we take $r_{\text{max}} = 100$, far beyond the range of observed interaction. We also implement the sampler in $\texttt{C}$ in an effort to manage the computational cost, e.g. by using pointers to avoid assignment of large arrays. For the largest datasets ($13000-14000$ points) the walltime is on the order of a few days.

\section{Applications to Data}\label{sec:application}
In this section we study the validity of our modeling and inferential approach, including our DMH algorithm, on simulated data. We then describe the application of our approach to the RSV data that motivated our work and draw scientific conclusions.

\subsection{Simulated Data}
Simulations with two different parameter settings were run as a means of testing the validity of inference obtained from the DMH algorithm. The first parameter setting attempts to emulate RSV-A data in that it imposes a fixed minimum allowable distance between infection marks, an artifact of the cell-staining procedure for this strain. The second attempts to emulate RSV-B data and does not impose a minimum distance between infected cells. At each parameter setting, three point patterns were independently simulated from our model and our inference approach was applied using DMH. The simulated point patterns consisted of $n \sim 3000$ points, for which inference took roughly 12 hours.

\begin{figure}[h]
\centering
\includegraphics[width = \textwidth]{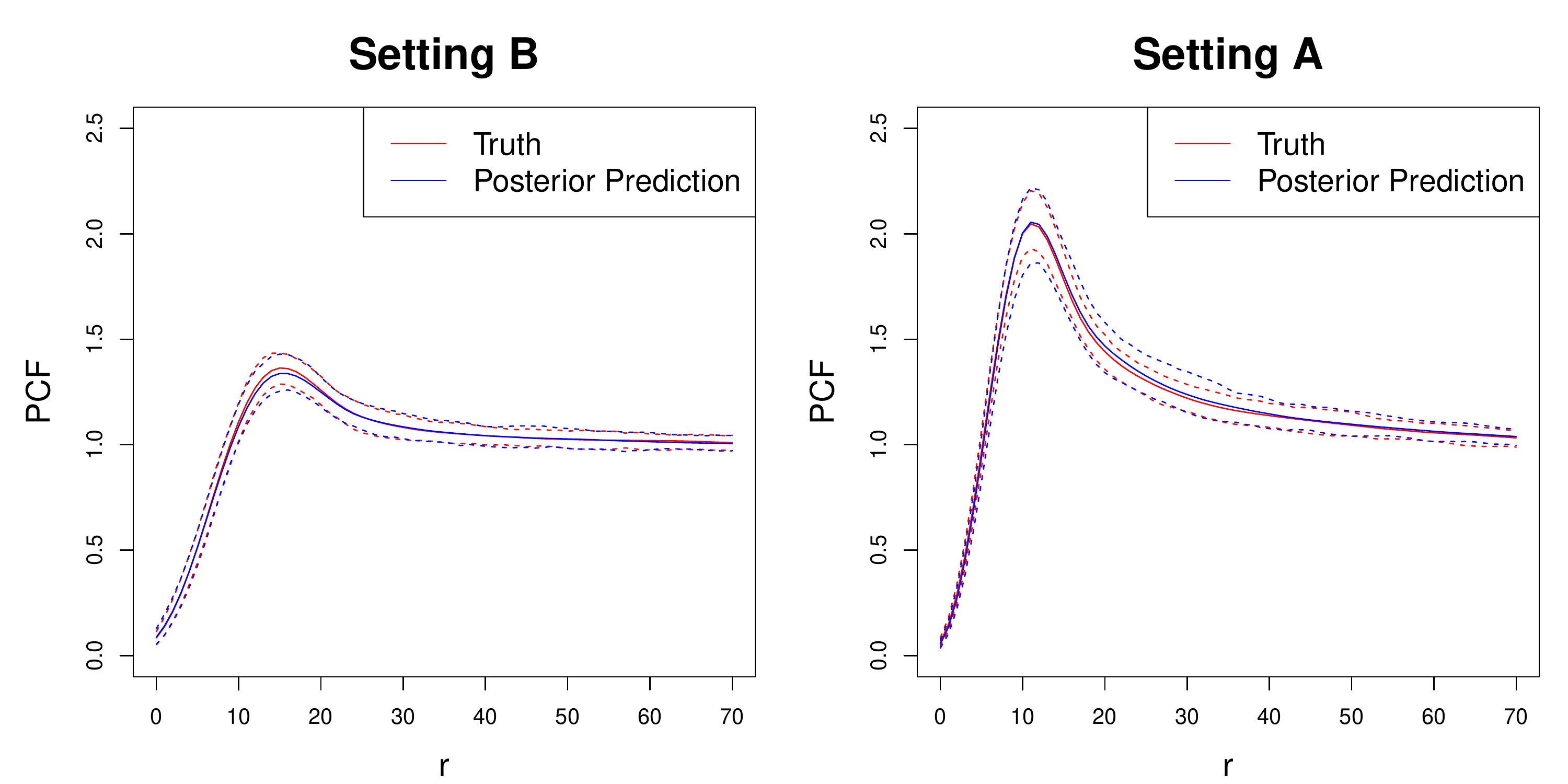}
\caption{Goodness-of-fit for simulated data. At both parameter settings we compare the PCF estimated from the true values of simulation parameters to a posterior predictive interval. The `true PCF' intervals are obtained by simulating 99 point patterns from the true parameters, computing the PCF of each one using the bootstrap method and taking 95\% pointwise intervals. The posterior predictive bands are obtained by simulating 99 point patterns from the parameters of our thinned Markov chain, computing the PCF of each one using the bootstrap method and taking 95\% pointwise intervals. If the true PCF falls within the posterior predictive interval then our model can faithfully reproduce the observed attraction behavior.}
\label{fig:PCFsim}
\end{figure}

We choose a gamma prior for $k$. For the remaining parameters our priors are uniform over a plausible range. To obtain 95\% credible intervals for the parameters, we use the highest posterior density (HPD) interval algorithm \citep{chen2000hpd}. Table~\ref{table:simRSV} demonstrates that in both settings the parameter values used to simulate the point patterns are recovered by our method of inference. In Figure \ref{fig:PCFsim} we also demonstrate goodness-of-fit for our simulations via the PCF.
\begin{table}
\caption{Simulations from the model for two different parameter settings. Setting 1 imposes a minimum allowable distance between points as is observed in the RSV-A data, while Setting 2 does not, as observed in the RSV-B data. In both cases the parameter values used to simulate the point patterns are recovered by DMH inference, lying within the 95\% HPD intervals. MCMC standard error via batch means \citep{jones2006,flegal2008} was below 0.01.}
\begin{tabular*}{\columnwidth}{@{}l@{\extracolsep{\fill}}c@{\extracolsep{\fill}}c@{\extracolsep{\fill}}c@{\extracolsep{\fill}}c@{\extracolsep{\fill}}c@{}}
\hline
Setting 1 & $\lambda \times 10^4$ & $\theta_1$ & $\theta_2$ & $\theta_3$ & $k$\\
\hline
Simulated Truth & 3 & 1.5 & 10 & 0.2 & 1.4 \\
Recovered Posterior Mean & 3.00  &  1.49  &  10.17  &  0.20  &  1.43 \\
95\% HPD Intervals & (2.85,3.16) & (1.46,1.53) & (9.84,10.51) & (0.18,0.22) & (1.32,1.54) \\
\hline
Setting 2 & $\lambda \times 10^4$ & $\theta_1$ & $\theta_2$ & $\theta_3$ & $k$\\
\hline
Simulated Truth & 4 & 1.2 & 15 & 0.3 & 1.2 \\
Recovered Posterior Mean & 4.02  &  1.19  &  15.18  &  0.31  &  2.06 \\
95\% HPD Intervals & (3.76,4.28) & (1.17,1.21) & (14.50,15.93) & (0.22,0.41) & (0.51,4.72) \\
\hline
\end{tabular*}
\label{table:simRSV}
\end{table}

\subsection{RSV Data}
Our spatial point process model is an effective tool to investigate the spatial patterns of infections in the RSV data. In particular, going beyond exploratory analyses, we can now formalize parametrically and make inference on the attraction-repulsion behavior of RSV-A or RSV-B infection marks for each of the experimental settings described in Section \ref{sec:setup}. This allows us to draw rigorous conclusions on whether and how, at various scales, infected cells tend to be closer together or farther apart than expected by chance. We can therefore gain insight into cell properties if we deduce how the susceptibility of a cell is affected by the presence of nearby infected cells. Our conclusions are based on inferences for the parameters of the interaction functions, which are reported in Table~\ref{table:RSV}.
\begin{table}
\centering
\caption{Posterior means and 95\% HPD intervals of model parameters for RSV-A and RSV-B. MCMC standard error via batch means \citep{jones2006,flegal2008} was below 0.01.}
\begin{tabular*}{\columnwidth}{@{}l@{\extracolsep{\fill}}c@{\extracolsep{\fill}}c@{\extracolsep{\fill}}c@{\extracolsep{\fill}}c@{\extracolsep{\fill}}c@{}}
\multicolumn{6}{c}{\centering \rule{0pt}{4ex} RSV-A}\\
\hline
& $\lambda \times 10^4$ & $\theta_1$ & $\theta_2$ & $\theta_3$ & $k$\\
\hline
\multicolumn{6}{c}{\centering 1A: single infection experiment}\\
\hline
3h & 3.78  &  1.29  &  11.95  &  0.16  &  1.42 \\
& (3.58,3.99) & (1.27,1.31) & (11.55,12.33) & (0.15,0.17) & (1.38,1.46) \\
16h & 5.27  &  1.22  &  13.39  &  0.18  &  1.41 \\
& (4.83,5.77) & (1.21,1.23) & (13.07,13.71) & (0.17,0.19) & (1.36,1.46) \\
\hline
\multicolumn{6}{c}{\centering 1B2A: RSV-A is the challenge to RSV-B}\\
\hline
3h & 3.79  &  1.31  &  10.17  &  0.15  &  1.28 \\
& (3.57,4.01) & (1.28,1.33) & (9.77,10.54) & (0.14,0.16) & (1.22,1.36) \\
16h & 3.48  &  1.44  &  9.02  &  0.159  &  0.85 \\
& (3.23,3.75) & (1.37,1.51) & (8.57,9.48) & (0.14,0.18) & (0.79,0.91) \\
\hline
\multicolumn{6}{c}{\centering 2A: RSV-A is the challenge, no primary infection}\\
\hline
3h & 3.58  &  1.33  &  10.67  &  0.16  &  1.20 \\
& (3.40,3.77) & (1.30,1.36) & (10.28,11.05) & (0.15,0.17) & (1.16,1.24) \\
16h & 3.23  &  1.46  &  9.51  &  0.18  &  1.55 \\
& (3.05,3.40) & (1.41,1.50) & (9.08,9.91) & (0.17,0.20) & (1.28,1.82) \\
\hline
\multicolumn{6}{c}{\centering \rule{0pt}{4ex} RSV-B}\\
\hline
 & $\lambda \times 10^4$ & $\theta_1$ & $\theta_2$ & $\theta_3$ & $k$\\
\hline
\multicolumn{6}{c}{\centering 1B: single infection experiment}\\
\hline
3h & 3.04  &  1.85  &  4.38  &  0.16  &  1.26 \\
& (2.90,3.19) & (1.76,1.93) & (4.06,4.70) & (0.15,0.17) & (1.21,1.31) \\
16h & 5.57  &  1.33  &  6.93  &  0.22  &  1.41 \\
& (5.24,5.93) & (1.31,1.36) & (6.63,7.30) & (0.21,0.24) & (1.36,1.45) \\
\hline
\multicolumn{6}{c}{\centering 1A2B: RSV-B is the challenge to RSV-A}\\
\hline
3h & 3.57  &  1.62  &  4.09  &  0.20  &  2.04 \\
& (3.36,3.78) & (1.55,1.70) & (3.65,4.48) & (0.18,0.23) & (1.63,2.43) \\
16h & 3.63  &  1.64  &  6.48  &  0.20  &  1.16 \\
& (3.41,3.87) & (1.59,1.70) & (6.14,6.83) & (0.18,0.22) & (1.11,1.20) \\
\hline
\multicolumn{6}{c}{\centering 2B: RSV-B is the challenge, no primary infection}\\
\hline
3h & 4.02  &  1.63  &  4.04  &  0.24  &  2.24 \\
& (3.80,4.26) & (1.56,1.71) & (3.66,4.47) & (0.21,0.27) & (1.90,2.57) \\
16h & 3.55  &  1.81  &  6.58  &  0.22  &  1.25 \\
& (3.37,3.77) & (1.76,1.87) & (6.29,6.85) & (0.20,0.23) & (1.21,1.30) \\
\hline
\end{tabular*}
\label{table:RSV}
\end{table}

\subsubsection{Conclusions}
\begin{figure}[p]
\centering
\includegraphics[width = \textwidth]{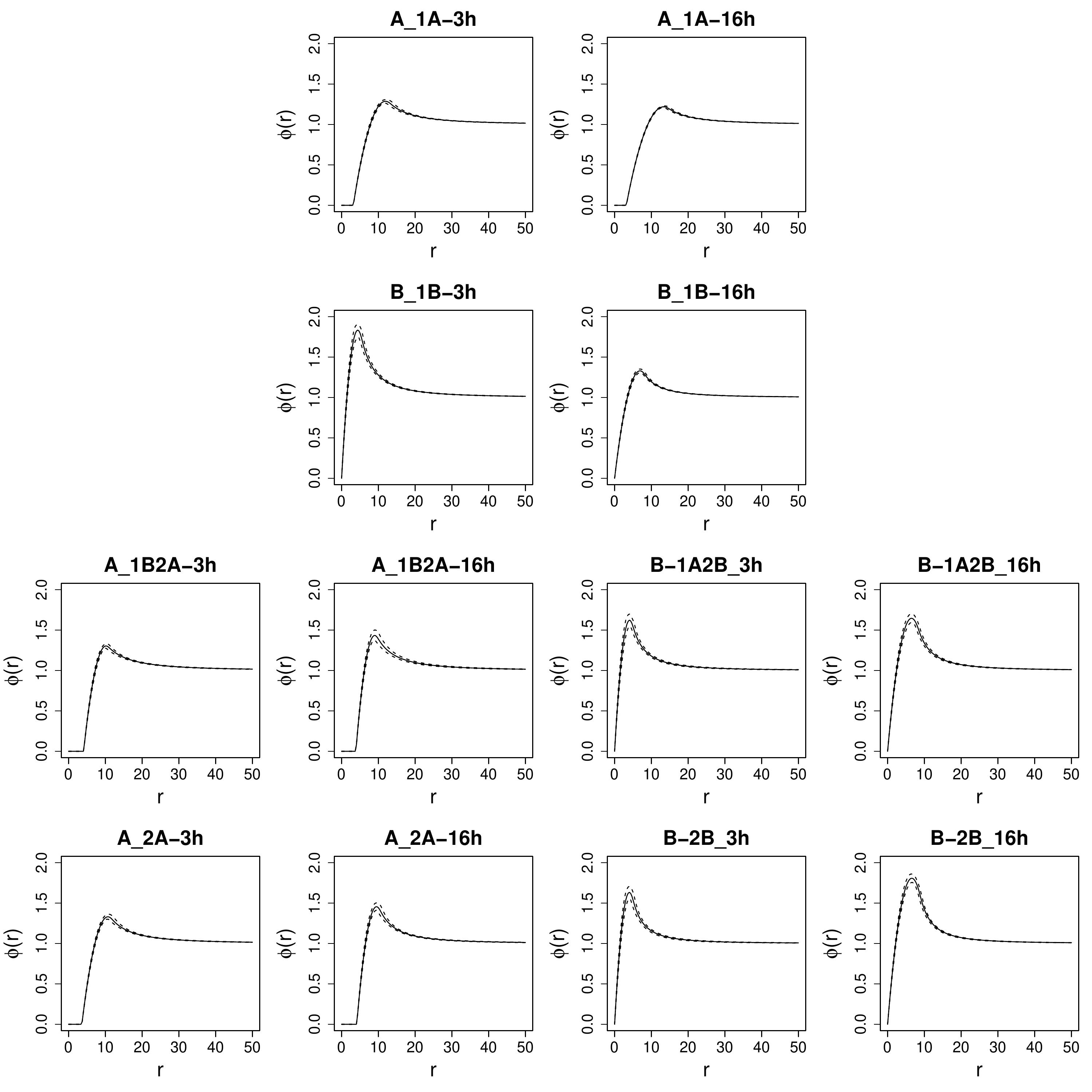}
\caption{Plots of inferred interaction functions with 95\% confidence intervals based on inferences on $\theta_1,\theta_2,\theta_3$ for each experimental setting.}
\label{fig:interactall}
\end{figure}

\begin{figure}[p]
\centering
\includegraphics[width = \textwidth]{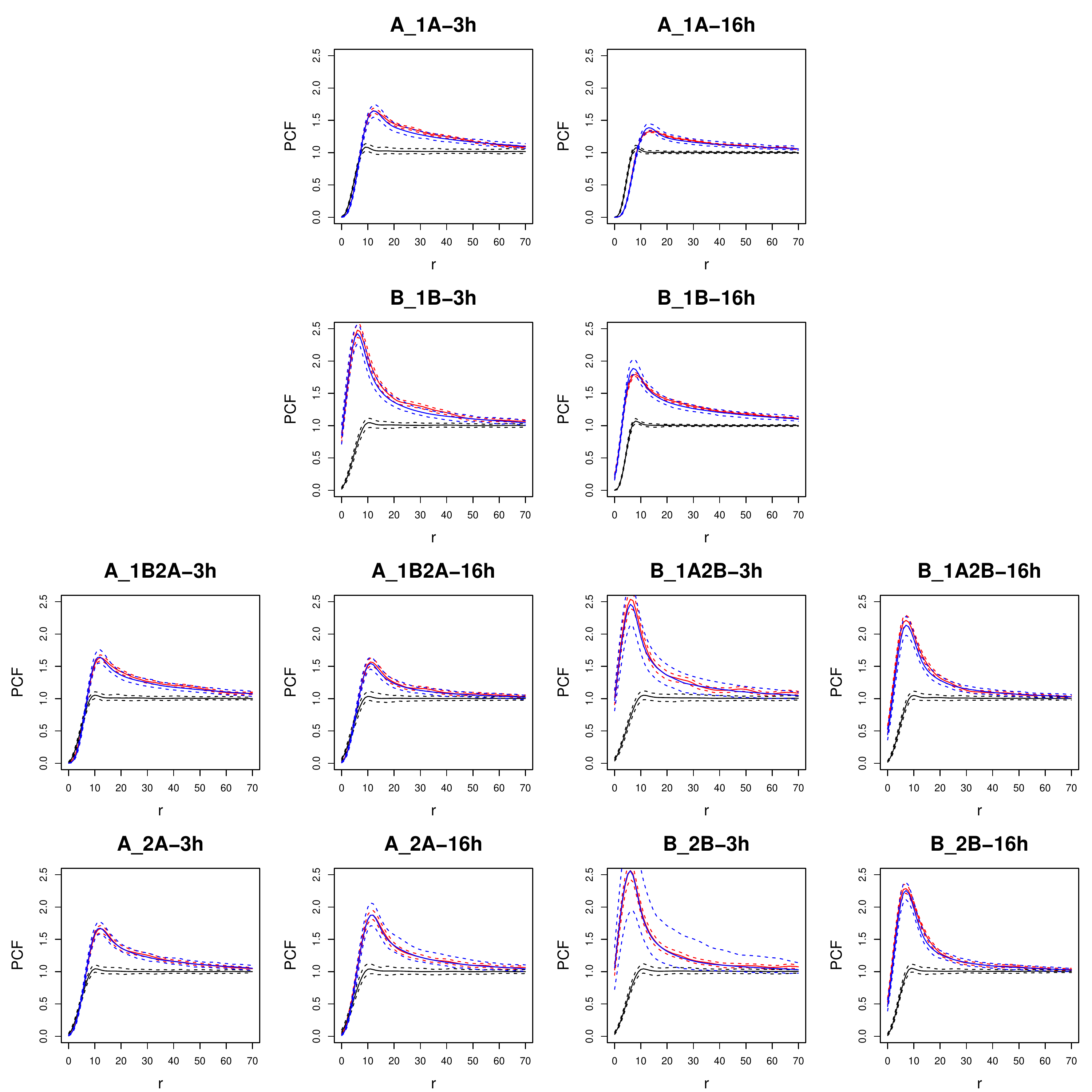}
\caption{Goodness-of-fit for all RSV experiments. To assess goodness-of-fit in our model we compare the PCF estimated from the data to a posterior predictive interval. The posterior predictive bands (blue) are obtained by simulating 99 point patterns from the parameters of our thinned Markov chain, computing the PCF of each one using the bootstrap method and taking 95\% pointwise intervals. If the empirical PCF (red) falls within the posterior predictive interval then our model can faithfully reproduce the observed attraction behavior. Null bands (black) are estimates of the PCF from {\it all} cells in the sample (as opposed to only infected cells) and provide a baseline for no attraction behavior.}
\label{fig:PCFall}
\end{figure}

\begin{itemize}
\item In all experimental settings, the peak $\theta_1$ is significantly greater than 1 (the 95\% HPD interval excludes 1). While the cell volume exclusion effect can cause repulsion at small scales, this is evidence of a significant attraction at scales further out. The propensity we observe for infected cells to lump together suggests, for both RSV-A and RSV-B, an increased susceptibility to infection for cells in the proximity of infected cells. This is not caused by the virus diffusing out from infected cells; there is not sufficient time in the experiment for this to happen, and moreover the virus is applied uniformly across the entire sample. Therefore, the fact that neighboring cells are infected is not indicative of ``spread'', but rather a synchronizing of cells across space indicative of communication between them.
\item All RSV-B experiments have a peak $(\theta_1)$ that is significantly larger and occur at a scale $(\theta_2)$ that is significantly smaller than their RSV-A counterparts. Thus, regardless of experimental conditions (e.g.~whether there was a primary infection, and the time lag between the primary and challenge infections) RSV-B infected cells have a higher propensity to lump together than RSV-A infected cells. This suggests that, when compared to RSV-A, RSV-B may induce a stronger increase in susceptibility to infection for cells neighboring infected cells.
\item When RSV-B is the challenge infection, in the absence of a primary infection the peak becomes significantly larger for longer time lags $\big( \hat{\theta}_{1,B_{2B-3h}} \in (1.56,1.71)$ vs. $\hat{\theta}_{1,B_{2B-16h}} \in (1.76,1.87) \big)$.
However, in the presence of a primary infection with RSV-A, the peak does not change significantly with the time lag $\big( \hat{\theta}_{1,B_{1A2B-3h}} \in (1.55,1.70)$ vs. $\hat{\theta}_{1,B_{1A2B-16h}} \in  (1.59,1.70) \big)$.
This is evidence that a primary infection with RSV-A induces a response that, given enough time, changes the susceptibility patterns to a subsequent challenge with RSV-B.
\item On the other hand, we observe no evidence that a primary infection with RSV-B induces an analogous response to RSV-A. When RSV-A is the challenge infection, there is a significant difference in the peak across time lags in both the absence
$\big( \hat{\theta}_{1,A_{2A-3h}} \in (1.30,1.36)$ vs. $\hat{\theta}_{1,A_{2A-16h}} \in (1.41,1.50) \big)$
and presence $\big( \hat{\theta}_{1,A_{1B2A-3h}} \in (1.28,1.33)$ vs. $\hat{\theta}_{1,A_{1B2A-16h}} \in (1.37,1.51) \big)$ of a primary infection by RSV-B.
\item In the single infection experiments, imaging infection marks after 3 or 16 hours result in significantly different spatial patterns. That is, the propensity for infected cells to lump together changes significantly depending on how long we wait. In RSV-A ($A_{1A-3h}$ and $A_{1A-16h}$) we see that the peak ($\theta_1$) is significantly lower and occurs at a distance ($\theta_2$) that is significantly greater after 16 hours than after 3 hours. A similar result is obtained for RSV-B ($B_{1B-3h}$ and $B_{1B-16h}$). This suggests that as an infection is given more time to act it  becomes more uniformly distributed across a cell culture.
\end{itemize}
Plots of inferred interaction functions for each experimental setting are included in Figure \ref{fig:interactall} along with a method of assessing goodness-of-fit for our model via the pair correlation function in Figure \ref{fig:PCFall}.

\section{Discussion}\label{sec:discussion}
The novel point process model we have developed lets us study in a rigorous way the complex and scale-dependent attraction-repulsion behavior of cells infected with RSV. In particular, our model allows for the strength of the interaction (attraction and repulsion) to vary smoothly with spatial scale; this flexibility is not available in existing models. In addition to providing a flexible approach for modeling attraction-repulsion behavior, our parametric specification of the interaction function allows us to draw meaningful and easy-to-interpret scientific conclusions based on parameter inference. Such interpretability is not readily available from existing approaches, for example nonparametric approaches like PCFs (discussed in our exploratory data analysis section). For instance, we are able to infer that exposure to RSV-A leads to an increased susceptibility of neighboring cells to infection with RSV-A. A similar result is found with RSV-B in response to an initial RSV-B exposure.

In previous work \citep{simeonov2010exploratory}, a somewhat qualitative version of these results were obtained by comparing K-statistics to simulated scenarios. It was therefore difficult to determine whether the response of RSV-A was similar to RSV-B and to answer such a question from the data.  Moving to a parametric framework allows us to not only determine whether the responses of the strains are different, but also {\it how} they are different.  Is it due to a slower spatial decay?  Is it due to a more rapid influence on neighboring cells?  Is the effect on susceptibility strongest at the width of one cell or the width of three cells?  The current methodology not only allow us to determine an effect, as was done previously, but to specify some aspects of the nature of the effect.

\begin{figure}[p]
\centering
\includegraphics[width = \textwidth]{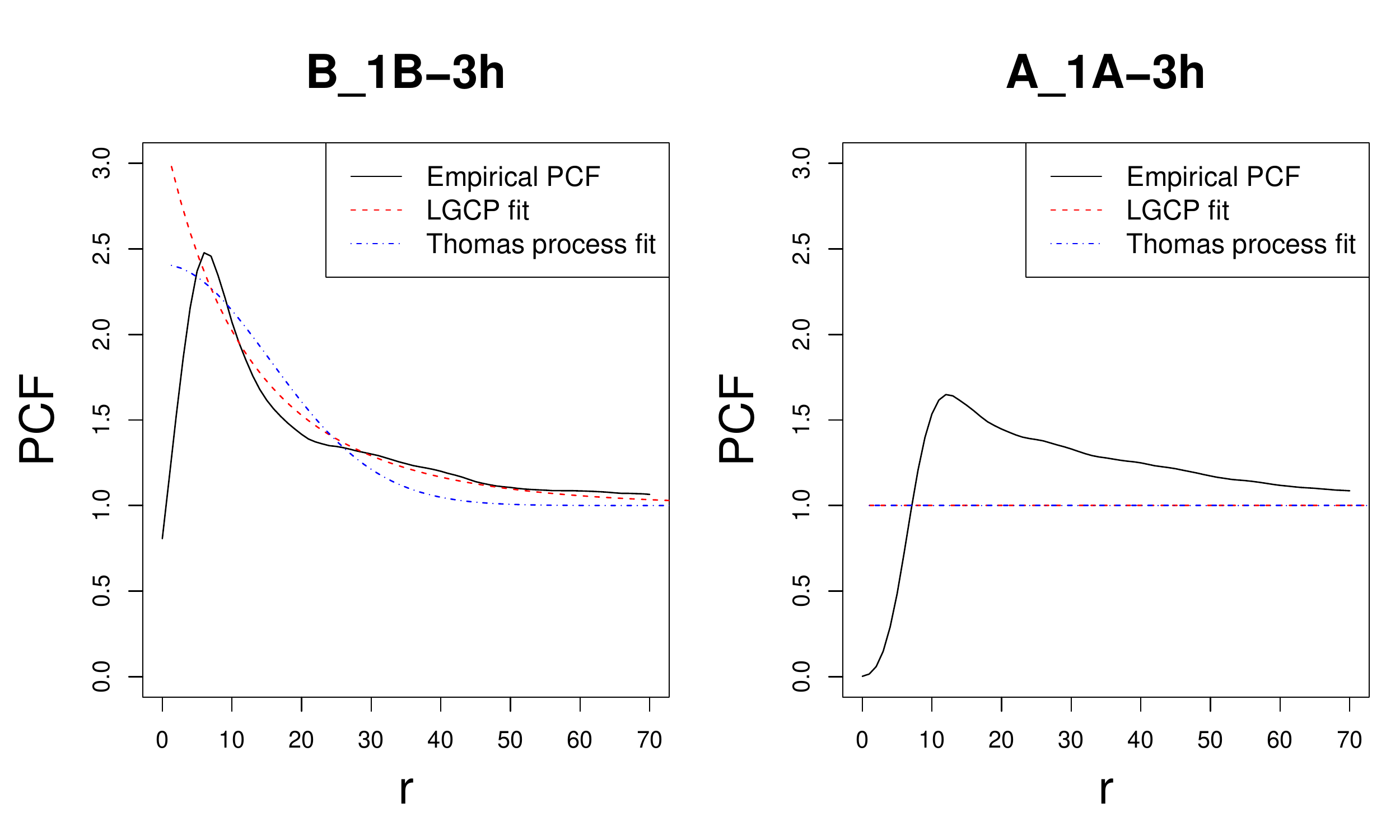}
\caption{Cox processes were fit to an RSV-B point pattern by the method of minimum contrast \citep{diggle1984} for the pair correlation function. The best fit PCFs for a Log Gaussian Cox Process and a Thomas Process are compared to the empirical PCF. We observe that while these models can reasonably capture the attraction behavior, they fail to capture the repulsion at small $r$. This problem is amplified for an RSV-A experiment with more repulsion; the best fit for both models reduces to $g(r)=1$. }
\label{fig:coxsim}
\end{figure}

We also attempted to capture the attraction and repulsion behavior with existing Cox process models. While these are not models for interactions between points one can indirectly get at the attraction behavior as we demonstrate in Figure \ref{fig:coxsim}. However, it was not clear how one could use the Cox process framework to readily capture both attraction and repulsion. For the scientific questions we investigate here it is important for us to use models that can capture both attraction and repulsion. If there were mechanisms that induce neighbors of infected cells to resist infections these would be of critical scientific interest. For instance, the question we have addressed indirectly here of whether primary infection with one strain decreases susceptibility to a challenge with another strain. We even have some preliminary evidence that decreased susceptibility may indeed be the case.

The work presented here could be extended in several ways. We have analyzed infection marks for RSV-A and RSV-B separately. For experimental settings in which both infections are present (one as the primary and one as the challenge) we could potentially use a marked point process model to formalize and analyze directly the spatial association between cells infected with RSV-A and RSV-B. This is problematic for our data setting because A/B stains cannot be uniquely identified with cell nuclei but we believe that such an extension may be pursued in cases where the data allow for it. On a different front, we have assessed the temporal progression of viral infections and related responses comparing parameter inferences across just two discrete time points (3h and 16h); given data on more time points, it would be very interesting to integrate our framework into a fully spatiotemporal model.

Our parametric approach is limited in that the parametric specification of the interaction function needs to be flexible enough to match the observed attraction-repulsion behavior. For more complicated behaviors we may require a more complicated parametric form. In principle a nonparametric specification of the interaction would allow for greater flexibility but this would come at the cost of much greater computational complexity and potential identifiability issues. Nonparametric approaches may also be useful for exploratory purposes and to suggest appropriate model specifications.

In terms of computational viability, we note that using our version of the DMH algorithm we were able to handle inference on RSV datasets comprising up to roughly 13,000 spatial locations. Point patterns of this size would be beyond the scope of many commonly used inferential methods, even for models simpler than the one we have developed here. Nevertheless, our approach is still computationally expensive due to the nature of the algorithm's two nested MCMC samplers. Hence there are computational challenges in extending our framework to even larger datasets; this is a potential subject for future research.

\section*{Acknowledgements}

We are grateful to Mary Poss and her laboratory at the Pennsylvania State University for making the RSV data available to us, providing advice regarding the scientific problem, and offering helpful comments on the manuscript.

\clearpage

\section*{Appendix}

While it seems counterintuitive at first, it is not generally true that PCF estimates at small distances should have wider intervals. Consider for instance a hardcore process, with e.g. a hardcore radius of 3. Then all estimates of the PCF will be identically zero for $r \leq 3$ and the intervals would have zero width. Likewise, if the probability of two points within distance $r \leq 3$ is very small but nonzero, then the variability in the estimated PCF will be small, even though there are few pairs of points separated by these distances. This is what happens at small values of $r$ in our data due to the cell volume exclusion effect.

\label{lastpage}

\begin{thebibliography}{}

\bibitem[\protect\citeauthoryear{Atchad{\'e} et~al.}{2008}]{atchade2008bayesian}
Atchad{\'e}, Y., Lartillot, N., and Robert, C. P. (2008). Bayesian computation for statistical models with intractable normalizing constants. {\it arXiv preprint arXiv:0804.3152.}

\bibitem[\protect\citeauthoryear{Besag}{1974}]{besag1974spatial}
Besag, J. (1974). Spatial interaction and the statistical analysis of lattice systems. {\it Journal of the Royal Statistical Society. Series B (Methodological)} {\bf 36}, 192--236.

\bibitem[\protect\citeauthoryear{Chen et~al.}{2000}]{chen2000hpd}
Chen, M-H, Shao, Q-M, and Ibrahim, J. G. (2000). {\it Monte Carlo methods in Bayesian computation.} New York: Springer.

\bibitem[\protect\citeauthoryear{Diggle}{1983}]{diggle1983statistical}
Diggle, P. J. (1983). {\it Statistical analysis of spatial point patterns.} London: Academic Press.

\bibitem[\protect\citeauthoryear{Diggle and Gratton}{1984}]{diggle1984}
Diggle, P. J., and Gratton, R. J. (1984). Monte Carlo methods of inference for implicit statistical models. {\it Journal of the Royal Statistical Society. Series B (Methodological)}, 193-227.

\bibitem[\protect\citeauthoryear{Flegal et~al.}{2008}]{flegal2008}
Flegal, J. M., Haran, M., and Jones, G. L. (2008). Markov chain Monte Carlo: Can we trust the third significant figure? {\it Statistical Science} {\bf 23}, 250--260.

\bibitem[\protect\citeauthoryear{Geyer}{1999}]{geyer1999likelihood}
Geyer, C. J. (1999). Likelihood inference for spatial point processes. {\it Stochastic geometry: likelihood and computation} {\bf 80}, 79-140.

\bibitem[\protect\citeauthoryear{Geyer and Thompson}{1992}]{geyer1992constrained}
Geyer, C. J., and Thompson, E. A. (1992). Constrained Monte Carlo maximum likelihood for dependent data. {\it Journal of the Royal Statistical Society. Series B (Methodological)} {\bf 54}, 657--699.

\bibitem[\protect\citeauthoryear{Jones et~al.}{2006}]{jones2006}
Jones, G. L., Haran, M., Caffo, B. S., and Neath, R. (2006). Fixed-width output analysis for Markov chain Monte Carlo. {\it Journal of the American Statistical Association} {\bf 101}, 1537--1547.

\bibitem[\protect\citeauthoryear{Kerscher et~al.}{2000}]{kerscher2000comparison}
Kerscher, M., Szapudi, I., and Szalay, A. S. (2000). A comparison of estimators for the two-point correlation function. {\it The Astrophysical Journal Letters} {\bf 535}, L13.

\bibitem[\protect\citeauthoryear{Liang}{2010}]{liang2010double}
Liang, F. (2010). A double Metropolis-Hastings sampler for spatial models with intractable normalizing constants. {\it Journal of Statistical Computation and Simulation} {\bf 80}, 1007--1022.

\bibitem[\protect\citeauthoryear{Loh}{2008}]{loh2008valid}
Loh, J. M. (2008). A valid and fast spatial bootstrap for correlation functions. {\it The Astrophysical Journal} {\bf 681}, 726.

\bibitem[\protect\citeauthoryear{M{\o}ller and Waagpetersen}{2003}]{moller2003statistical}
M{\o}ller, J. and Waagepetersen, R. P. (2003). {\it Statistical inference and simulation for spatial point processes.} CRC Press.

\bibitem[\protect\citeauthoryear{M{\o}ller et ~al.}{2004}]{moller2004efficient}
M{\o}ller, J., Pettitt, A. N., Reeves, R., and Berthelsen, K. K. (2004). An efficient Markov chain Monte Carlo method for distributions with intractable normalising constants. Research Report R-2004-02, Department of Mathematical Sciences, Aalborg University.

\bibitem[\protect\citeauthoryear{M{\o}ller et~al.}{2006}]{moller2006efficient}
M{\o}ller, J., Pettitt, A. N., Reeves, R., and Berthelsen, K. K. (2006). An efficient Markov chain Monte Carlo method for distributions with intractable normalising constants. {\it Biometrika} {\bf 93}, 451--458.

\bibitem[\protect\citeauthoryear{Murray et~al.}{2006}]{murray2006}
Murray, I., Gharhamani, Z., and MacKay, D. (2006). MCMC for doubly-intractable distributions, in {\it Proceedings of the 22nd Annual Conference on Uncertainty in Artificial Intelligence.} Arlington: AUAI Press.

\bibitem[\protect\citeauthoryear{Ripley}{1993}]{ripley1991statistical}
Ripley, B. D. (1991). {\it Statistical inference for spatial processes.} Cambridge University Press.

\bibitem[\protect\citeauthoryear{Simeonov}{2012}]{simeonov2012}
Simeonov, I. (2012). An application of spatial point process methods to RSV infection experiments. Master's Thesis, The Pennsylvania State University.

\bibitem[\protect\citeauthoryear{Simeonov et~al.}{2010}]{simeonov2010exploratory}
Simeonov, I., Gong, X., Kim, O., Poss, M., Chiaromonte, F., and Fricks, J. (2010). Exploratory spatial analysis of {\it in vitro} respiratory syncytial virus co-infections. {\it Viruses} {\bf 2}, 2782--2802.

\bibitem[\protect\citeauthoryear{Stoyan and Stoyan}{1994}]{stoyan1994fractals}
Stoyan, D., and Stoyan, H. (1994). {\it Fractals, random shapes, and point fields: methods of geometrical statistics.} Chichester: Wiley.

\bibitem[\protect\citeauthoryear{Strauss}{1975}]{strauss1975model}
Strauss, D. J. (1975). A model for clustering. {\it Biometrika} {\bf 62}, 467--475.

\bibitem[\protect\citeauthoryear{Wang and Landau}{2001}]{wang2001efficient}
Wang, F., and Landau, D. P. (2001). Efficient, multiple-range random walk algorithm to calculate the density of states. {\it Physical Review Letters} {\bf 86}, 205.
\end{thebibliography}
\end{document}